\documentclass[epj]{webofc}
\usepackage[varg]{txfonts}   % Web of Conferences font
\wocname{EPJ Web of Conferences}
\woctitle{Confinement 2016}
\newcommand{\be}{\begin{equation}}
\newcommand{\ee}{\end{equation}}

\begin{document}
\selectlanguage{english}
\title{Chiral magnetic superconductivity}
%
% subtitle (optional, strongly discouraged)

\author{Dmitri E. Kharzeev\inst{1,2,3}\fnsep\thanks{\email{dmitri.kharzeev@stonybrook.edu}}}
% \and
%        Second author\inst{2} \and
%        Third author\inst{3}
%        % etc.
%}

\institute{Department of Physics and Astronomy, Stony Brook University,
Stony Brook, New York 11794-3800, USA
\and
           Department of Physics, Brookhaven National Laboratory, Upton, NY 11973, USA
\and
     RIKEN-BNL Research Center, Brookhaven National Laboratory,
Upton, New York 11973-5000, USA      
}

\abstract{%
  Materials with charged chiral quasiparticles in external parallel electric and magnetic fields can support an electric current that grows linearly in time, corresponding to diverging DC conductivity. From experimental viewpoint, this ``Chiral Magnetic Superconductivity" (CMS) is thus analogous to conventional superconductivity. However the underlying physics is entirely different -- the CMS does not require a condensate of Cooper pairs breaking the gauge degeneracy, and is thus not accompanied by Meissner effect. Instead, it owes its existence to the (temperature-independent) quantum chiral anomaly and the conservation of chirality.  As a result, this phenomenon can be expected to survive to much higher temperatures. Even though the chirality of quasiparticles is not strictly conserved in real materials, the chiral magnetic superconductivity should still exhibit itself in AC measurements at frequencies larger than the chirality-flipping rate, and in microstructures of Dirac and Weyl semimetals with thickness below the mean chirality-flipping length that is about $1-100\ \mu$m. In nuclear physics, the CMS should contribute to the charge-dependent elliptic flow in heavy ion collisions. 
}
\maketitle
\section{Introduction}
\label{intro}
The chiral magnetic effect (CME) is the generation of electric current induced by an external magnetic field in the presence of imbalance between the left- and right-handed chiral fermions. The CME and related phenomena are the subject of vigorous ongoing research in particle, nuclear, and condensed matter physics, and I will not attempt to cover it here -- reviews can be found in Refs. \cite{Kharzeev:2013ffa,Kharzeev:2012ph,Kharzeev:2015znc}.

Instead, I would like to address the relation between the CME and conventional superconductivity (SC). Both are non-dissipative transport phenomena, but as we will see the underlying physics is drastically different. Nevertheless, the 
CME and SC do share some common properties, and accentuating the similarities and the differences between these macroscopic quantum phenomena may be useful for focusing the future research in this area.

\section{London theory of superconductivity}
\label{sec-1}

Let us start with a brief recap of ``conventional" superconductivity. Instead of microscopic BCS theory introducing the condensate of Cooper pairs, we will base our discussion on the approach proposed in 1935 by Fritz and Heinz London \cite{London1935}. The London equations describe electromagnetic response of a superconductor and can be derived from the following bold assumption about the proportionality between the electric current ${\bf J}$ and the vector gauge potential {\bf A}:
\be\label{london1}
{\bf J} = - \mu^2 {\bf A},  
\ee
where a phenomenological constant $\mu$ is related to the density of superconducting carriers $n_s$, electron mass $m$ and electric charge $e$ by $\mu = (e^2\ n_s/m)^{1/2}$.  

This equation is striking, as it is in manifest conflict with gauge invariance. Indeed, the electric current is a measurable physical quantity, and is thus gauge-invariant. On the other hand, the gauge potential obviously changes under 
gauge transformations, ${\bf A} \to {\bf A} + {\bf \nabla} \varphi$, where $\varphi$ is an arbitrary scalar function. At present we understand that this {\it apparent} breaking of gauge invariance is due to the Higgs-Anderson mechanism and is triggered by the formation of Cooper pair condensate in the ground state of a superconductor; it is in fact perfectly consistent with the gauge symmetry, as we will discuss below. Nevertheless, to apply the equation (\ref{london1})
we need to specify a gauge condition, and following Londons we will use the Coulomb gauge ${\bf \nabla} {\bf A} = 0$. 

In Coulomb gauge, the electric field is ${\bf E} = - {\bf \dot A}$ and the magnetic field is ${\bf B} = {\bf \nabla \times \bf A}$; therefore, the assumption (\ref{london1}) yields the following two equations of superconductor electrodynamics:
\be\label{lon_el}
{\bf \dot J} = \mu^2\ {\bf E},
\ee
and 
\be\label{lon_mag}
{\bf \nabla} \times {\bf J} = - \mu^2\ {\bf B}.
\ee
The first of these equations (\ref{lon_el}) tells us that a constant external electric field ${\bf E}$ generates inside a superconductor a transient state with an electric current that grows linearly in time.  Once the electric field is turned off (or screened away), Eq. (\ref{lon_el}) tells us that the current ${\bf J}$ is not allowed to decay and should persist -- this is superconductivity.

For our forthcoming discussion of the relation between superconductivity (SC) and chiral magnetic effect (CME), it is instructive to look at (\ref{london1}) from the viewpoint of fundamental discrete symmetries. Under parity transformation P, both vectors {\bf J} and {\bf A} flip sign, so the quantity $\mu^2$ is P-even, unlike the CME conductivity \cite{Fukushima:2008xe,Kharzeev:2009pj} defined by 
\be\label{cme}
{\bf J} = \frac{e^2}{2 \pi^2}\ \mu_5\ {\bf B} \equiv 
\sigma_{CME}\ {\bf B}
\ee
that is P-odd ($e$ is electric charge and $\mu_5 = (\mu_L - \mu_R)/2$ is the chiral chemical potential describing the difference of chemical potentials for left- and right-handed fermions). 

Under time reversal T, both vectors {\bf J} and {\bf A} again flip sign, and so $\mu^2$ is even under T, just like the CME conductivity. There is however a very essential difference between SC and CME -- in the latter case, since ${\bf B}$ is a gauge-invariant quantity, the absence of dissipation does not require breaking gauge degeneracy, and can thus be achieved without any condensates in the ground state. Since condensates are usually destroyed by thermal fluctuations, this makes CME potentially a more robust phenomenon that can survive to much higher temperatures than SC.

Being even under time reversal is a highly unusual property for a conductivity -- for example, the Ohmic conductivity defined by ${\bf J} = \sigma\ {\bf E}$ is T-odd (the electric current ${\bf J}$ is T-odd, but electric field ${\bf E}$ is T-even). This is natural, because the Ohmic conductivity describes the processes of dissipation, and dissipation leads to the increase of entropy. The growth of entropy according to the second law of thermodynamics cannot be reversed, and thus generates an arrow of time. On the other hand, a conductivity that is T-even describes the processes that can be reversed in time, and thus have to be non-dissipative \cite{Kharzeev:2011ds}. 

To gain a deeper insight, we can consider the power ${\rm P}$ dissipated by Ohmic current:
\be
{\rm P_{Ohm}} = \int d^3 x\ {\bf J} \cdot {\bf E} = \sigma \int d^3 x\ {\bf E}^2 .
\ee
The quantity on the r.h.s. is positive-definite, and is proportional to the energy of electric field. Therefore, the dissipated power has to be positive as well. 

On the other hand, for superconductors we get
\be\label{power_sc}
{\rm P_{SC}} = \int d^3 x\ {\bf J} \cdot {\bf E} = - \mu^2 \int d^3 x\ {\bf A \cdot E} =  \frac{\mu^2}{2} \frac{d}{dt} \int d^3 x\ {\bf A}^2 .
\ee
Once the electric field is switched off, ${\bf E} =0$, eq. (\ref{lon_el}) tells us that the current persists, but the dissipated power according to (\ref{power_sc}) vanishes. 

The quantity ${\bf A}^2$ on the r.h.s. of (\ref{power_sc}) appears to break gauge invariance, but this is only apparent: as is well known, the action of a superconductor is formulated in terms of the combination $A_\mu - 1/e\ \partial_\mu \phi$, where $\phi$ is the phase of the condensate field. In the ground state of a superconductor,   
${\bf A} = 1/e\ {\bf \nabla} \phi$ is a pure gauge, so there is thus no magnetic field, ${\bf B} = {\bf \nabla} \times {\bf A} = 0$. On the other hand, a deviation from the ground state with ${\bf B} \neq 0$ requires ${\bf A}$ proportional to a coordinate transverse to the direction of ${\bf B}$, so the corresponding contribution to the free energy $\int d^3 x\ {\bf A}^2$ grows faster than the volume  (note that the value of this integral is minimal in the Coulomb gauge \cite{Gubarev:2000eu}). The magnetic field is thus energetically unfavorable and is expelled by the superconductor  -- this is the Meissner effect. Therefore both the superconductivity and the Meissner effect can be seen as a direct consequence of (\ref{london1}), and its symmetry properties.

\section{Chiral magnetic effect as a new type of superconductivity}

Let us now consider the corresponding expression for CME:
\be\label{cme_power}
{\rm P_{CME}} = \int d^3 x\ {\bf J} \cdot {\bf E} = \sigma_{\rm CME} \int d^3 x\ {\bf E \cdot B} = - \frac{1}{2} \sigma_{\rm CME} \frac{d}{dt} \int d^3 x\ {\bf A \cdot B} .
\ee
It is proportional to the time derivative of magnetic helicity $h_m \equiv \int d^3 x\ {\bf A \cdot B}$ that is a topological Chern-Simons 3-form describing the topology of magnetic flux. This invariant counts the chirality of a knot formed by the lines of magnetic field in a given gauge configuration.  The expression (\ref{cme_power}) thus shows that as long as the topology of magnetic flux does not change, the power dissipated by the chiral magnetic current is equal to zero. 
This demonstrates that the CME current is protected by topology of the gauge field. In fact, a topological mechanism for one-dimensional superconductivity was introduced long time ago by Frohlich \cite{Frohlich1954}, see \cite{Wiegmann,Abanov} for more recent developments.

The chiral magnetic instability \cite{Joyce:1997uy,Boyarsky:2011uy,Boyarsky:2015faa,Tashiro:2012mf,Hirono:2015rla,Buividovich:2015jfa,Yamamoto:2016xtu,Tuchin:2016qww,Gorbar:2016klv} can result in the transfer of chirality carried by fermions into magnetic helicity, but this decay happens at rather long space and time scales -- in fact, this mechanism may be responsible for the generation of magnetic fields in the Universe. At late times, the instability leads to a self-similar cascade towards the Chandrasekhar-Kendall states that minimize magnetic energy for a given value of magnetic helicity \cite{Hirono:2015rla,Yamamoto:2016xtu}. In a condensed matter setup this decay can be eliminated, if the characteristic instability wavelength is smaller than the size of sample \cite{Hirono:2015rla}. 
If the decay does occur, the CME current can dissipate into magnetic helicity, as illustrated by the expression (\ref{cme_power}).

To make an analogy with superconductivity, let us consider a chiral material (i.e. a material that hosts massless charged chiral fermions, such as a Dirac or Weyl semimetal, or a quark-gluon plasma) in external parallel magnetic and electric fields. First, let us assume that the chirality of  fermions is strictly conserved. 
The chiral anomaly of quantum electrodynamics \cite{Adler:1969gk,Bell:1969ts} then dictates that the electric and magnetic fields generate the chiral charge density $\rho_5$ with the rate given by
\be\label{anom_eq}
\frac{d \rho_5}{dt} = \frac{e^2}{4 \pi^2 \hbar^2 c} \bf{E}\cdot \bf{B} ,
\ee
and the chiral charge density grows linearly in time: 
\be
\rho_5 = \frac{e^2}{4 \pi^2 \hbar^2 c} {\bf{E}\cdot \bf{B}} \ t .
\ee
Let us now relate the density of the chiral charge $\rho_5$ and the chiral chemical potential $\mu_5$ by introducing $\chi \equiv \partial \rho_5  / \partial \mu_5$, so that  $\rho_5 = \chi \mu_5 + ...$ and $\mu_5 \simeq  \chi^{-1} \rho_5$ for small $\mu_5$; note that at late times when $\mu_5$ grows larger than other scales in the system, this relation will eventually be violated. 

Using (\ref{cme}) for ${\bf E} || {\bf B}$ we now get
\be\label{cme_sup}
{\bf J} = \frac{e^2}{2 \pi^2}\ \mu_5\ {\bf B} = \frac{e^4}{8 \pi^4 \hbar^2 c}\ \chi^{-1} {\rm B}^2\ {\bf E}\ t \equiv \mu_{\rm CME}^2\ {\bf E}\ t.
\ee
Taking the time derivative on both sides, we get
\be\label{lon_cme}
{\bf \dot J} = \mu_{\rm CME}^2\ {\bf E},
\ee
which is completely analogous to the corresponding expression for superconductivity (\ref{lon_el})! We thus see that the CME in an ideal chiral material can be viewed as a new type of superconductivity -- we will call it Chiral Magnetic Superconductivity (CMS). In particular, if the electric field is switched off, the CME current according to (\ref{lon_cme}) persists. In real experiments, the current will eventually decay, but on a time scale set by the chirality-flipping transitions that is much longer than a typical transport time.

%%%%%%%

Despite the similarity between SC and CMS discussed above, the physics underlying the conventional and chiral magnetic superconductivities is entirely different: in the case of CMS, the relation (\ref{cme}) tells us that no breaking of gauge degeneracy is needed, in contrast to superconductivity (\ref{london1}). Because of this, there is no Meissner effect -- in the free energy instead of $\int d^3 x\ {\bf A}^2$, we now have a gauge-invariant quantity $\int d^3 x\ {\bf B}^2$ that scales with the volume, so the magnetic field is not expelled. 

On the contrary, the CMS is driven by magnetic field, and the corresponding conductivity in weak magnetic fields is proportional to the square of magnetic field strength, see (\ref{cme_sup}). In strong magnetic fields, when $B \gg T, \mu$ (where $T$ is the temperature and $\mu$ is the chemical potential), the fermions are frozen on the lowest Landau level, and $\rho_5$ and $\chi$ become proportional to $eB/2\pi$, the density of Landau levels in the transverse plane. In this case, the relation (\ref{cme_sup}) tells us that the CMS current should become proportional to ${\rm B}$ instead of ${\rm B}^2$.

In real materials, the chirality of fermions is not conserved exactly, and can be flipped through chirality-flipping scattering, or due to the surface Fermi arcs in the case of Weyl semimetals. Introducing the characteristic chirality-flipping time $\tau_V$, we then note that the CMS is limited to the AC response at frequencies $\omega >  \tau_V^{-1}$. Indeed, in the presence of chirality-flipping transitions, one should add the chirality loss term to the r.h.s. of the anomaly equation (\ref{anom_eq}): 
\be\label{anomalyeq}
\frac{d \rho_5}{dt} = \frac{e^2}{4 \pi^2 \hbar^2 c} \bf{E}\cdot \bf{B} - \frac{\rho_5}{\tau_V} .
\ee
At short times $t < \tau_V$, the corresponding CME current will still grow linearly in time. However at late times
$t \gg \tau_V$ the system approaches a stationary state with the chiral charge density
\be
\rho_5 = \frac{e^2}{4 \pi^2 \hbar^2 c} \bf{E}\cdot \bf{B} \ \tau_V .
\ee
The corresponding CME current when ${\bf E} || {\bf B}$ is given by
\be
{\bf J} = \frac{e^4}{8 \pi^4 \hbar^2 c}\ \chi^{-1} {\rm B}^2 \ \tau_V\ {\bf E} .
\ee
When paired with a conventional Ohmic current, the CME current induces a negative magnetoresistance \cite{Son:2012bg,Burkov} with a characteristic Lorentzian dependence on magnetic field. 
This CME current has recently been observed \cite{Li:2014bha,Xiong2015,Li2015,Huang2015} through the negative magneto-resistance in Dirac and Weyl semimetals (see also \cite{Kim2013} for a result in a topological insulator).

\section{Chiral magnetic conductivity in microstructures}

Since many potential applications of CME will probably rely on the AC current, it is worth to emphasize again that at frequencies above  $\omega_\chi \equiv \tau_V^{-1}$ the CME response is close to superconducting. The chiral magnetic conductivity at finite frequency drops \cite{Kharzeev:2009pj}, and magnetization current has to be taken into account to describe the AC response properly \cite{Kharzeev:2016sut}. Unlike at zero frequency, the AC response is characterized by a finite dissipation that grows with frequency \cite{Kharzeev:2009pj}, so to stay close to non-dissipative regime we need to 
choose frequency not much larger than $\omega_\chi$. 

In currently available chiral materials, the frequency $\omega_\chi$ is probably in the range of 10 GHz to 1 THz, see e.g. \cite{Li:2014bha,Xiong2015}; the corresponding chirality flipping time is between $10^{-12}$ s and $10^{-10}$ s. During this time, the fermion quasiparticles propagate without chirality-flipping backscattering over the distances that in ballistic regime are on the order of $l_\chi \simeq v_F \tau_V$, where $v_F$ is the quasiparticle Fermi velocity (in the diffusive regime, the relevant distance is $l_\chi \sim (D \tau_V)^{1/2}$, where $D$ is the diffusion constant). Assuming $\tau_V = 0.001 - 0.1$ ns, and $v_F = 1/300\ c$, we estimate that chiral fermions can preserve their chirality up to rather long distances of $l_\chi \sim 1 - 100\ {\rm \mu m}$. The propagation of the chiral charge through the material can be detected using the ``chiral battery" \cite{Fukushima:2008xe,Kharzeev:2012dc} -- in an external magnetic field, the chiral charge due to the CME induces an electric voltage. This phenomenon can be studied using the nonlocal transport measurement in which the chiral charge pumped by parallel electric and magnetic fields between one pair of terminals is detected through the electric voltage induced between the other pair of terminals located a distance $L$ away \cite{Parameswaran2013}. 

A recent experiment \cite{Zhang2015} has observed this phenomenon in ${\rm Cd_3 As_2}$, and used the dependence $\sim \exp(-L/l_\chi)$ of nonlocal transport on the separation $L$ between the terminals to extract the characteristic chirality-flippling length $l_\chi$. The corresponding value $l_\chi \simeq 2\ {\rm \mu m}$ is on the lower side of the range estimated above; however, it is still much larger than the typical mean free path of charge carriers. Remarkably, the value of $l_\chi$ appears almost independent of temperature up to $T \simeq 300$ K \cite{Zhang2015} -- this observation suggests that chirality is robust even at room temperature. Previously, similar results were obtained for 2D graphene \cite{Abanin2011}, so the approximate conservation of chirality appears a common feature for both 2D and 3D chiral materials -- and in the latter case, it enables anomaly-induced transport. 
\vskip0.3cm

The conservation of chirality over long distances suggests another path towards achieving CMS in 3D chiral materials -- instead of using the finite frequency response at $\omega > \omega_\chi$, one can use microstructures with thickness $l < l_\chi \sim 1-100 \ {\rm \mu m}$. Such microstructures are still thick enough to support 3D chiral quasiparticles, but are not too thick to induce chirality flips through the bulk scattering. In this case, we expect that even a response to constant electric and magnetic fields can be superconducting. 
\vskip0.3cm 

This conclusion has an interesting implication for CME in the quark-gluon plasma, since the ``samples" produced in heavy ion collisions are of size comparable to $l_\chi$. The colliding ions produce strong electromagnetic fields \cite{Kharzeev:2007jp},  and electric and magnetic fields are, on the average, parallel above the reaction plane, and antiparallel below it. The chiral magnetic superconductivity would thus contribute to the electric quadrupole deformation manifested in the charge-dependent elliptic flow and predicted as a signature of the chiral magnetic wave \cite{Kharzeev:2010gd,Burnier:2011bf,Gorbar:2011ya} and observed at RHIC \cite{Adamczyk:2015eqo} and LHC \cite{Adam:2015vje}.
Detailed estimates of these phenomena will be presented elsewhere.

\vskip0.3cm
I am indebted to Q. Li and V.I. Zakharov for valuable and stimulating discussions. 
This work was supported by in part by the U.S.
Department of Energy under Contracts No. DE-FG- 88ER40388 and DE-AC02-98CH10886.


\begin{thebibliography}{}
%\cite{Kharzeev:2013ffa}
\bibitem{Kharzeev:2013ffa} 
  D.~E.~Kharzeev,
  %``The Chiral Magnetic Effect and Anomaly-Induced Transport,''
  Prog.\ Part.\ Nucl.\ Phys.\  {\bf 75}, 133 (2014)
  doi:10.1016/j.ppnp.2014.01.002
  [arXiv:1312.3348 [hep-ph]].
  %%CITATION = doi:10.1016/j.ppnp.2014.01.002;%%
  %104 citations counted in INSPIRE as of 14 Dec 2016

%\cite{Kharzeev:2012ph}
\bibitem{Kharzeev:2012ph} 
  D.~E.~Kharzeev, K.~Landsteiner, A.~Schmitt and H.~U.~Yee,
  %``'Strongly interacting matter in magnetic fields': an overview,''
  Lect.\ Notes Phys.\  {\bf 871}, 1 (2013)
  doi:10.1007/978-3-642-37305-3
  [arXiv:1211.6245 [hep-ph]].
  %%CITATION = doi:10.1007/978-3-642-37305-3_1;%%
  %125 citations counted in INSPIRE as of 15 Dec 2016

%\cite{Kharzeev:2015znc}
\bibitem{Kharzeev:2015znc} 
  D.~E.~Kharzeev, J.~Liao, S.~A.~Voloshin and G.~Wang,
  %``Chiral magnetic and vortical effects in high-energy nuclear collisions?A status report,''
  Prog.\ Part.\ Nucl.\ Phys.\  {\bf 88}, 1 (2016)
  doi:10.1016/j.ppnp.2016.01.001
  [arXiv:1511.04050 [hep-ph]].
  %%CITATION = doi:10.1016/j.ppnp.2016.01.001;%%
  %73 citations counted in INSPIRE as of 14 Dec 2016
  
  \bibitem{London1935}
  F. London and H. London, Proc. Royal Society A: Mathematical, Physical and Engineering Sciences, {\bf 149}, 71 (1935). 

%\cite{Fukushima:2008xe}
\bibitem{Fukushima:2008xe} 
  K.~Fukushima, D.~E.~Kharzeev and H.~J.~Warringa,
  %``The Chiral Magnetic Effect,''
  Phys.\ Rev.\ D {\bf 78}, 074033 (2008)
  doi:10.1103/PhysRevD.78.074033
  [arXiv:0808.3382 [hep-ph]].
  %%CITATION = doi:10.1103/PhysRevD.78.074033;%%
  %765 citations counted in INSPIRE as of 14 Dec 2016
  
  %\cite{Kharzeev:2009pj}
\bibitem{Kharzeev:2009pj} 
  D.~E.~Kharzeev and H.~J.~Warringa,
  %``Chiral Magnetic conductivity,''
  Phys.\ Rev.\ D {\bf 80}, 034028 (2009)
  doi:10.1103/PhysRevD.80.034028
  [arXiv:0907.5007 [hep-ph]].
  %%CITATION = doi:10.1103/PhysRevD.80.034028;%%
  %155 citations counted in INSPIRE as of 14 Dec 2016
  
  %\cite{Kharzeev:2011ds}
\bibitem{Kharzeev:2011ds} 
  D.~E.~Kharzeev and H.~U.~Yee,
  %``Anomalies and time reversal invariance in relativistic hydrodynamics: the second order and higher dimensional formulations,''
  Phys.\ Rev.\ D {\bf 84}, 045025 (2011)
  doi:10.1103/PhysRevD.84.045025
  [arXiv:1105.6360 [hep-th]].
  %%CITATION = doi:10.1103/PhysRevD.84.045025;%%
  %83 citations counted in INSPIRE as of 14 Dec 2016
  
  %\cite{Adler:1969gk}
\bibitem{Adler:1969gk} 
  S.~L.~Adler,
  %``Axial vector vertex in spinor electrodynamics,''
  Phys.\ Rev.\  {\bf 177}, 2426 (1969).
  doi:10.1103/PhysRev.177.2426
  %%CITATION = doi:10.1103/PhysRev.177.2426;%%
  %3381 citations counted in INSPIRE as of 14 Dec 2016
  
  %\cite{Bell:1969ts}
\bibitem{Bell:1969ts} 
  J.~S.~Bell and R.~Jackiw,
  %``A PCAC puzzle: pi0 --> gamma gamma in the sigma model,''
  Nuovo Cim.\ A {\bf 60}, 47 (1969).
  doi:10.1007/BF02823296
  %%CITATION = doi:10.1007/BF02823296;%%
  %2759 citations counted in INSPIRE as of 14 Dec 2016
  
  %\cite{Gubarev:2000eu}
\bibitem{Gubarev:2000eu} 
  F.~V.~Gubarev, L.~Stodolsky and V.~I.~Zakharov,
  %``On the significance of the vector potential squared,''
  Phys.\ Rev.\ Lett.\  {\bf 86}, 2220 (2001)
  doi:10.1103/PhysRevLett.86.2220
  [hep-ph/0010057].
  %%CITATION = doi:10.1103/PhysRevLett.86.2220;%%
  %195 citations counted in INSPIRE as of 15 Dec 2016
  
    \bibitem{Frohlich1954}
  G. Frohlich, Proc. R. Soc. London {\bf A 223}, 296 (1954).
  
  \bibitem{Wiegmann}
  P.B. Wiegmann, Prog. Theor. Phys. Suppl. {\bf 107}, 243 (1992).
  
  \bibitem{Abanov}
  A.G. Abanov and P.B. Wiegmann, Phys. Rev. Letters {\bf 86}, 1319 (2001). 
  
  %\cite{Joyce:1997uy}
\bibitem{Joyce:1997uy} 
  M.~Joyce and M.~E.~Shaposhnikov,
  %``Primordial magnetic fields, right-handed electrons, and the Abelian anomaly,''
  Phys.\ Rev.\ Lett.\  {\bf 79}, 1193 (1997)
  doi:10.1103/PhysRevLett.79.1193
  [astro-ph/9703005].
  %%CITATION = doi:10.1103/PhysRevLett.79.1193;%%
  %199 citations counted in INSPIRE as of 15 Dec 2016
  
  %\cite{Boyarsky:2011uy}
\bibitem{Boyarsky:2011uy} 
  A.~Boyarsky, J.~Frohlich and O.~Ruchayskiy,
  %``Self-consistent evolution of magnetic fields and chiral asymmetry in the early Universe,''
  Phys.\ Rev.\ Lett.\  {\bf 108}, 031301 (2012)
  doi:10.1103/PhysRevLett.108.031301
  [arXiv:1109.3350 [astro-ph.CO]].
  %%CITATION = doi:10.1103/PhysRevLett.108.031301;%%
  %71 citations counted in INSPIRE as of 15 Dec 2016
 
  
  %\cite{Boyarsky:2015faa}
\bibitem{Boyarsky:2015faa} 
  A.~Boyarsky, J.~Frohlich and O.~Ruchayskiy,
  %``Magnetohydrodynamics of Chiral Relativistic Fluids,''
  Phys.\ Rev.\ D {\bf 92}, no. 4, 043004 (2015)
  [arXiv:1504.04854 [hep-ph]].
  %%CITATION = ARXIV:1504.04854;%%
  %3 citations counted in INSPIRE as of 18 Oct 2015

%\cite{Tashiro:2012mf}
\bibitem{Tashiro:2012mf} 
  H.~Tashiro, T.~Vachaspati and A.~Vilenkin,
  %``Chiral Effects and Cosmic Magnetic Fields,''
  Phys.\ Rev.\ D {\bf 86}, 105033 (2012)
  doi:10.1103/PhysRevD.86.105033
  [arXiv:1206.5549 [astro-ph.CO]].
  %%CITATION = doi:10.1103/PhysRevD.86.105033;%%
  %41 citations counted in INSPIRE as of 15 Dec 2016

%\cite{Hirono:2015rla}
\bibitem{Hirono:2015rla} 
  Y.~Hirono, D.~Kharzeev and Y.~Yin,
  %``Self-similar inverse cascade of magnetic helicity driven by the chiral anomaly,''
  Phys.\ Rev.\ D {\bf 92}, no. 12, 125031 (2015)
  doi:10.1103/PhysRevD.92.125031
  [arXiv:1509.07790 [hep-th]].
  %%CITATION = doi:10.1103/PhysRevD.92.125031;%%
  %18 citations counted in INSPIRE as of 15 Dec 2016

%\cite{Buividovich:2015jfa}
\bibitem{Buividovich:2015jfa} 
  P.~V.~Buividovich and M.~V.~Ulybyshev,
  %``Numerical study of chiral plasma instability within the classical statistical field theory approach,''
  Phys.\ Rev.\ D {\bf 94}, no. 2, 025009 (2016)
  doi:10.1103/PhysRevD.94.025009
  [arXiv:1509.02076 [hep-th]].
  %%CITATION = doi:10.1103/PhysRevD.94.025009;%%
  %10 citations counted in INSPIRE as of 15 Dec 2016

%\cite{Yamamoto:2016xtu}
\bibitem{Yamamoto:2016xtu} 
  N.~Yamamoto,
  %``Scaling laws in chiral hydrodynamic turbulence,''
  Phys.\ Rev.\ D {\bf 93}, no. 12, 125016 (2016)
  doi:10.1103/PhysRevD.93.125016
  [arXiv:1603.08864 [hep-th]].
  %%CITATION = doi:10.1103/PhysRevD.93.125016;%%
  %6 citations counted in INSPIRE as of 15 Dec 2016

%\cite{Tuchin:2016qww}
\bibitem{Tuchin:2016qww} 
  K.~Tuchin,
  %``Spontaneous topological transitions of electromagnetic fields in spatially inhomogeneous CP-odd domains,''
  arXiv:1607.07481 [hep-ph].
  %%CITATION = ARXIV:1607.07481;%%

%\cite{Gorbar:2016klv}
\bibitem{Gorbar:2016klv} 
  E.~V.~Gorbar, I.~Rudenok, I.~A.~Shovkovy and S.~Vilchinskii,
  %``Anomaly-driven inverse cascade and inhomogeneities in a magnetized chiral plasma in the early Universe,''
  Phys.\ Rev.\ D {\bf 94}, no. 10, 103528 (2016)
  doi:10.1103/PhysRevD.94.103528
  [arXiv:1610.01214 [hep-ph]].
  %%CITATION = doi:10.1103/PhysRevD.94.103528;%%
  
    %\cite{Son:2012bg}
\bibitem{Son:2012bg} 
  D.~T.~Son and B.~Z.~Spivak,
  %``Chiral Anomaly and Classical Negative Magnetoresistance of Weyl Metals,''
  Phys.\ Rev.\ B {\bf 88}, 104412 (2013)
  doi:10.1103/PhysRevB.88.104412
  [arXiv:1206.1627 [cond-mat.mes-hall]].
  %%CITATION = doi:10.1103/PhysRevB.88.104412;%%
  %117 citations counted in INSPIRE as of 15 Dec 2016
  
  \bibitem{Burkov}
  A. A. Burkov, Phys. Rev. Letters, {\bf 113}, 247203 (2014); 
  
  doi:https://doi.org/10.1103/PhysRevLett.113.247203
  
  %\cite{Li:2014bha}
\bibitem{Li:2014bha} 
  Q. Li,	D. E. Kharzeev, 	C. Zhang,	Y. Huang,	I. Pletikosic, A. V. Fedorov,	R. D. Zhong,	J. A. Schneeloch,	G. D. Gu and T. Valla, 
  %``Observation of the chiral magnetic effect in ZrTe5,''
  Nature Phys.\  {\bf 12}, 550 (2016)
  doi:10.1038/nphys3648
  [arXiv:1412.6543 [cond-mat.str-el]].
  %%CITATION = doi:10.1038/nphys3648;%%
  %83 citations counted in INSPIRE as of 15 Dec 2016
  
\bibitem{Xiong2015}
J. Xiong, S. K. Kushwaha, T. Liang, J. W. Krizan, M. Hirschberger, W. Wang, R. J. Cava, N. P. Ong, 
Science\ {\bf 350}, 413 (2015).
  doi: 10.1126/science.aac6089
  
  \bibitem{Li2015}
  C.-Z. Li, L.-X. Wang, H. Liu, J. Wang, Z.-M. Liao and D.-P. Yu, 
  Nature Comm. {\bf 6},  (2015).
doi:10.1038/ncomms10137

\bibitem{Huang2015}
X. Huang, L. Zhao, Y. Long, P. Wang, D. Chen, Z. Yang, H. Liang, M. Xue, H. Weng, Z. Fang, Xi Dai, and G. Chen, 
Physical
Review X {\bf 5}, 031023 (2015).

\bibitem{Kim2013}
H.-J. Kim, K.-S. Kim, J. F. Wang, M. Sasaki, N. Satoh, A. Ohnishi, M. Kitaura, M. Yang, L. Li, 
 Phys. 
Rev. Letters {\bf 111}, 246603 (2013).

  %\cite{Kharzeev:2016sut}
\bibitem{Kharzeev:2016sut} 
  D.~E.~Kharzeev, M.~A.~Stephanov and H.~U.~Yee,
  %``Anatomy of chiral magnetic effect in and out of equilibrium,''
  arXiv:1612.01674 [hep-ph].
  %%CITATION = ARXIV:1612.01674;%%
  
  %\cite{Kharzeev:2012dc}
\bibitem{Kharzeev:2012dc} 
  D.~E.~Kharzeev and H.~U.~Yee,
  %``Anomaly induced chiral magnetic current in a Weyl semimetal: Chiral electronics,''
  Phys.\ Rev.\ B {\bf 88}, no. 11, 115119 (2013)
  doi:10.1103/PhysRevB.88.115119
  [arXiv:1207.0477 [cond-mat.mes-hall]].
  %%CITATION = doi:10.1103/PhysRevB.88.115119;%%
  %19 citations counted in INSPIRE as of 15 Dec 2016
  
  \bibitem{Parameswaran2013}
  S.A. Parameswaran, T. Grover, D. A. Abanin, D. A. Pesin, A. Vishwanath, 
  Phys. Rev. X {\bf 4}, 031035 (2014).
  doi:https://doi.org/10.1103/PhysRevX.4.031035

 \bibitem{Zhang2015}
  C. Zhang {\it et al.}, arXiv:1504.07698; {\it to appear in} Nature Comm.
  %%CITATION = ARXIV:1504.07698;%%
  
  \bibitem{Abanin2011}
  D. A. Abanin, S. V. Morozov, L. A. Ponomarenko, R. V. Gorbachev, A. S. Mayorov, M. I. Katsnelson, K. Watanabe, T. Taniguchi, K. S. Novoselov, L. S. Levitov, A. K. Geim, 
  Science {\bf 332}, 328 (2011). 
  doi:	10.1126/science.1199595
  
  %\cite{Kharzeev:2007jp}
\bibitem{Kharzeev:2007jp} 
  D.~E.~Kharzeev, L.~D.~McLerran and H.~J.~Warringa,
  %``The Effects of topological charge change in heavy ion collisions: 'Event by event P and CP violation',''
  Nucl.\ Phys.\ A {\bf 803}, 227 (2008)
  doi:10.1016/j.nuclphysa.2008.02.298
  [arXiv:0711.0950 [hep-ph]].
  %%CITATION = doi:10.1016/j.nuclphysa.2008.02.298;%%
  %831 citations counted in INSPIRE as of 15 Dec 2016
  
  %\cite{Kharzeev:2010gd}
\bibitem{Kharzeev:2010gd} 
  D.~E.~Kharzeev and H.~U.~Yee,
  %``Chiral Magnetic Wave,''
  Phys.\ Rev.\ D {\bf 83}, 085007 (2011)
  doi:10.1103/PhysRevD.83.085007
  [arXiv:1012.6026 [hep-th]].
  %%CITATION = doi:10.1103/PhysRevD.83.085007;%%
  %153 citations counted in INSPIRE as of 15 Dec 2016
  
  %\cite{Burnier:2011bf}
\bibitem{Burnier:2011bf} 
  Y.~Burnier, D.~E.~Kharzeev, J.~Liao and H.~U.~Yee,
  %``Chiral magnetic wave at finite baryon density and the electric quadrupole moment of quark-gluon plasma in heavy ion collisions,''
  Phys.\ Rev.\ Lett.\  {\bf 107}, 052303 (2011)
  doi:10.1103/PhysRevLett.107.052303
  [arXiv:1103.1307 [hep-ph]].
  %%CITATION = doi:10.1103/PhysRevLett.107.052303;%%
  %121 citations counted in INSPIRE as of 15 Dec 2016
  
  %\cite{Gorbar:2011ya}
\bibitem{Gorbar:2011ya} 
  E.~V.~Gorbar, V.~A.~Miransky and I.~A.~Shovkovy,
  %``Normal ground state of dense relativistic matter in a magnetic field,''
  Phys.\ Rev.\ D {\bf 83}, 085003 (2011)
  doi:10.1103/PhysRevD.83.085003
  [arXiv:1101.4954 [hep-ph]].
  %%CITATION = doi:10.1103/PhysRevD.83.085003;%%
  %51 citations counted in INSPIRE as of 15 Dec 2016
  
  \bibitem{Adamczyk:2015eqo} 
  L.~Adamczyk {\it et al.} [STAR Collaboration],
  ``Observation of charge asymmetry dependence of pion elliptic flow and 
  the possible chiral magnetic wave in heavy-ion collisions,''
  Phys.\ Rev.\ Lett.\  {\bf 114}, no. 25, 252302 (2015)
  [arXiv:1504.02175 [nucl-ex]].
 
\bibitem{Adam:2015vje} 
  J.~Adam {\it et al.} [ALICE Collaboration],
  ``Charge-dependent flow and the search for the chiral magnetic wave 
  in Pb-Pb collisions at $\sqrt{s_{\rm NN}} =$ 2.76 TeV,''
  Phys.\ Rev.\ C {\bf 93}, no. 4, 044903 (2016)
  [arXiv:1512.05739 [nucl-ex]].
  
\end{thebibliography}
\end{document}